\documentclass[12pt]{article}
\usepackage[english]{babel}
\usepackage[utf8x]{inputenc}
\usepackage[T1]{fontenc}
\usepackage{wrapfig}
\usepackage{cite}
\usepackage{ae,aecompl}
\usepackage{enumitem}
\usepackage{bm}
\usepackage{amsmath}
\usepackage{amssymb}
\usepackage{graphicx}
\usepackage{multicol}
\usepackage[colorinlistoftodos]{todonotes}
\usepackage[colorlinks=true, allcolors=blue]{hyperref}
\usepackage[letterpaper,top=1in,bottom=1in,left=1in,right=1in,marginparwidth=1in]{geometry}
\usepackage{hyperref}

\vspace{-5cm}
\title{{Astro2020 Science White Paper}\\
\vspace{0.7cm}
\bf{Cosmological Probes of Dark Matter Interactions: The Next Decade}}
\date{}
\author{}

\begin{document}
\maketitle
\vspace{-2cm}
\begin{center}
\textbf{Thematic area:} Cosmology and Fundamental Physics.     
\end{center}

\noindent\textbf{Principal Author:} \\
Vera Gluscevic\\
Institution: University of Florida\\
Email: \texttt{v.gluscevic@ufl.edu}\\
Phone: 1-352-392-8754

\vspace{0.5cm}
\noindent\textbf{Co-Authors:}

\noindent 
{Yacine Ali-Ha\"imoud$^{1}$, Keith Bechtol$^{2}$, Kimberly K.~Boddy$^{3}$, C\'eline B\oe{}hm$^{4,5}$, Jens Chluba$^{6}$, Francis-Yan Cyr-Racine$^{7,8}$, Cora Dvorkin$^{8}$, Vera Gluscevic$^{9,10}$, Daniel Grin$^{11}$, Julien Lesgourgues$^{12}$, Mathew S. Madhavacheril$^{13}$, Samuel D.~McDermott$^{14}$, Julian B.~Mu\~noz$^{8}$, Ethan O.~Nadler$^{15}$, Vivian Poulin$^{16,3}$, Sarah Shandera$^{17}$, Katelin Schutz$^{18}$, Tracy R.~Slatyer$^{19,20}$, Benjamin Wallisch$^{20,21}$}

\def\affil#1{\noindent #1 \\}

\begin{multicols}{2}
\scriptsize
\affil{$^{1}$ CCPP, Department of Physics, New York University}
\affil{$^{2}$ Department of Physics, University of Wisconsin-Madison}
\affil{$^{3}$ Department of Physics and Astronomy, Johns Hopkins Univ.}
\affil{$^{4}$ School of Physics, The University of Sydney, Australia}
\affil{$^{5}$ LAPTH, France; Perimeter institute, Canada}
\affil{$^{6}$ Jodrell Bank Center for Astrophysics, UK}
\affil{$^{7}$ Department of Physics and Astronomy, Univ.~of New Mexico}
\affil{$^{8}$ Department of Physics, Harvard University}
\affil{$^{9}$ Department of Physics, University of Florida}
\affil{$^{10}$ Department of Physics, Princeton University}
\affil{$^{11}$ Department of Physics and Astronomy, Haverford College}
\affil{$^{12}$ RWTH Aachen University, Germany}
\affil{$^{13}$ Department of Astrophysical Sciences, Princeton University}
\affil{$^{14}$ Fermi National Accelerator Laboratory}
\affil{$^{15}$ KIPAC and Department of Physics, Stanford University}
\affil{$^{16}$ LUPM, CNRS \& Universit\'e de Montpellier, France}
\affil{$^{17}$ Department of Physics, Pennsylvania State University}
\affil{$^{18}$ Department of Physics, University of California Berkeley}
\affil{$^{19}$ Center for Theoretical Physics, MIT}
\affil{$^{20}$ School of Natural Sciences, Institute for Advanced Study}
\affil{$^{21}$ Department of Physics, University of California San Diego}

\normalsize
\end{multicols}

{\small \noindent\textbf{Endorsers:} The full list of names is available at:\\\url{https://github.com/veragluscevic/Astro2020-DM-Cosmology-Endorsers}}
\begin{abstract}
Cosmological observations offer unique and robust avenues for probing the fundamental nature of dark matter particles---they broadly test a range of compelling theoretical scenarios, often surpassing or complementing the reach of terrestrial and other experiments.
We discuss observational and theoretical advancements that will play a pivotal role in realizing a strong program of cosmological searches for the identity of dark matter in the coming decade. Specifically, we focus on measurements of the cosmic-microwave-background anisotropy and spectral distortions, and tracers of structure (such as the Lyman-$\alpha$ forest, galaxies, and the cosmological 21-cm signal).
\end{abstract}

\pagebreak
\section{Key Question: What is Dark Matter?}
 
Observations of the Universe, from our Galactic neighborhood to the cosmological horizon, consistently testify that $\sim$85$\%$ of matter behaves as a cold non-collisional fluid that sources gravitational potentials and underpins structure on virtually all observable scales.
\textbf{Over the past decades, it has been confidently established that the main constituents of the dark matter (DM) component \textit{cannot} be any known baryonic particles.}
The existence of DM thus implies new physics whose investigation centrally drives research at the intersection of modern astrophysics, cosmology, and particle physics. 

A versatile range of laboratory experiments has been built worldwide to directly detect or produce some of the best-motivated particle candidates that could account for cosmological DM: WIMPs, WIMP-like particles, axions, etc; so far, they have no positive result.
Astrophysical and cosmological observations provide the only evidence for the existence of DM and source a large portion of what is known about its properties: its stability on cosmological time scales, its apparent non-collisional nature, and its central role in the formation and growth of structure.
Recently, those observations have also emerged as a powerful probe of DM microphysics, complementary in reach to laboratory experiments.

\begin{wrapfigure}{r}{0.5\textwidth}
\begin{center}
\vspace{-0.9cm}
\includegraphics[width=0.5\textwidth]{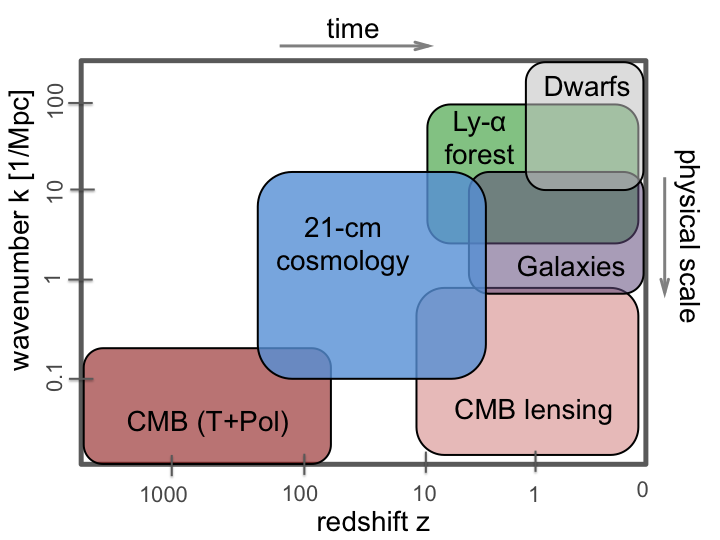}
\end{center}
\vspace{-0.8cm}
\caption{Approximate cosmological epochs and physical scales corresponding to different observables (spectral distortions are omitted for compactness of presentation; their origin lies to the left of the plot).}
\vspace{-0.2cm}
\label{fig:scales}
\end{wrapfigure}
We focus on cosmological searches, whose key goal is to detect the effects of DM interactions on the structure and thermal history of the Universe, and use them to pin down the particle identity of DM.
\textbf{Cosmology is a versatile tool that can test broad classes of theoretical scenarios}; here, we focus on DM interactions with known particles (baryons, photons, and neutrinos), interactions within the DM sector itself, and DM annihilations and decays.\footnote{We note that other science white papers focus on scenarios in which DM consists of ultra-light axion-like particles, warm DM, primordial black holes, etc.}
\textbf{The next decade of observations will see a tremendous leap in sensitivity to each of them}, in some cases, by many orders of magnitude in DM mass and interaction cross sections.
Multiple observables---cosmic microwave background (CMB) anisotropy and spectral distortions, tracers of large-scale structure, and objects in our Galactic neighborhood---can all probe the same underlying microphysics throughout cosmic history, on a broad range of physical scales (Fig.~\ref{fig:scales}).
\textbf{Establishing a DM signal discovery and robustly testing its fundamental nature will rely on confirmation and consistency checks between different data sets.}
This will necessitate progress in understanding synergies between observables, and development of frameworks that enable their joint analyses to probe the same underlying DM microphysics.
In addition, the DM search program with next-generation cosmological data will rely on accurate theoretical treatments of the formation and evolution of structure in non-standard cosmologies, and modeling of baryonic and other non-linear effects that present systematic uncertainties for DM parameter inference \cite{2019arXiv190201055D,2018PhRvD..98h3540M,AliHaimoud_19}.
We summarize the future promise of cosmological measurements in search for the identity of DM, and conclude with a list of advancements that will be pivotal to their success in the coming decade.
\vspace{-0.4cm}
\section{Theory and Observations}
\label{sec:thobs}
\vspace{-0.2cm}
\subsection{Scattering with Baryons}

In the standard model of cosmology DM is non-collisional, but elastic scattering between DM and visible particles is commonplace in some of the best-motivated DM models, including WIMPs and WIMP-like particles.
For this reason, terrestrial DM searches with direct detection experiments are looking for scattering of Galactic-halo DM on nuclei in underground targets.
However, the same scattering processes can occur in a cosmological setting and lead to an exchange of heat and momentum between DM and baryons (i.e.,the visible particles), absent in the standard (non-collisional) cosmology \cite{Boddy:2018kfv,Gluscevic:2017ywp,Boddy:2018wzy,Xu:2018efh,Slatyer:2018aqg,Dvorkin:2013cea,2001PhLB..518....8B,2005A&A...438..419B,2004NuPhB.683..219B,Sigurdson:2004zp}. 
This can affect both the thermal history and the evolution of cosmological perturbations, captured by various observables.

\textbf{CMB spectral distortions.} 
DM can drain heat from the primordial plasma, by scattering with either protons, electrons, or photons in the early Universe \cite{AliHaimoud_15}.
Heat exchange occurring later than two months after the Big Bang cannot fully thermalize and thus leads to distortions of the CMB frequency spectrum away from a perfect blackbody \cite{Hu_96,2012MNRAS.419.1294C}.  
The CMB frequency spectrum is most sensitive to light particles that thermally decouple as early as $z \sim 2 \times 10^6$.
The existing COBE FIRAS measurements provide upper limits on DM-baryon and DM-photon interactions for DM masses lower than $\sim 0.1$ MeV. 
These limits are similarly stringent but independent of those derived from CMB anisotropy from \textit{Planck}.
Future measurements that can detect a fractional distortion of order $\sim 10^{-8}$ would be sensitive to interacting DM particles as massive as $\sim 1$ GeV \cite{AliHaimoud_15}. 
Spectral distortions in conjunction with power suppression in CMB anisotropy in future data could yield robust evidence for DM physics taking place in the very early Universe.

\textbf{CMB anisotropy.} 
DM-baryon scattering prior to recombination ($z\sim 1100$) induces a drag force between the DM and baryon fluids, smoothing density fluctuations more prominently at progressively smaller scales and for stronger interactions.
As perturbations grow, the lack of small-scale structure translates into an underabundance of luminous DM tracers throughout cosmic history (various tracers are illustrated in Fig.~\ref{fig:scales}).

\textit{Planck} measurements of the CMB temperature anisotropy currently provide the most pristine cosmological bound on the DM-proton scattering cross section~\cite{Boddy:2018kfv,Gluscevic:2017ywp,Boddy:2018wzy,Xu:2018efh,Slatyer:2018aqg,Dvorkin:2013cea}.
They already sensitively probe DM particles with masses outside the detection limits of most existing direct detection experiments ($\gtrsim 1$ keV) \cite{Gluscevic:2017ywp}, through their interactions with baryons when the Universe was only a thousand years old.
However, the first high-signal-to-noise measurements of CMB polarization and lensing on $\sim$arcmin angular scales---to be delivered by the next-generation CMB experiments---enable a leap in sensitivity to DM scattering cross sections by up to \textit{several orders of magnitude} beyond \textit{Planck}~\cite{2018PhRvD..98l3524L,2019JCAP...02..056A,Abazajian:2016yjj,2019arXiv190210541H}.
Interestingly, the effects of DM-baryon interactions are distinct from other new physics targeted by future CMB experiments (e.g., the neutrino mass, etc.), and their measurement is insensitive to uncertainties on cosmological parameters.
CMB anisotropy is therefore a robust probe of DM, that can be further refined with application of high-accuracy treatments of DM signals \cite{AliHaimoud_19}.
However, in spite of its robustness, the CMB has a limited grasp of small scales, where the effects of the interactions are more prominent. 
Small scales are accessible to other observables, but are more prone to systematic modeling and measurement uncertainties.

\textbf{Lyman-${\boldsymbol\alpha}$ forest.} 
For example, SDSS Lyman-$\alpha$ forest measurements trace the matter power spectrum on comoving scale of about 1 Mpc; their analysis set some of the most stringent to-date cosmological bounds on DM-baryon interactions \cite{Xu:2018efh,Dvorkin:2013cga}.
These bounds imply that a proton residing in a Milky-Way-like galaxy does \textit{not} scatter with DM over the age of the galaxy.
Future spectroscopic surveys could improve upon these limits by reconstructing even smaller scales with similar measurements.
However, inference of fundamental physics from these data relies on accurate modeling of the Lyman-$\alpha$ forest and non-linear evolution of small scales \cite{2017PhRvD..96b3522I}.
Future joint analyses with high-precision CMB measurements and probes that have orthogonal systematic uncertainties could alleviate such issues.

\textbf{21-cm cosmology.} 
Over the coming decade, measurements of the hyperfine 21-cm line in atomic hydrogen from the cosmic Dark Ages and Reionization will likely start to probe redshifts $10 \lesssim z \lesssim 200$, far beyond galaxy surveys.
The strength of the 21-cm signal is proportional to the difference between the temperature of the gas (i.e., baryons) and of the CMB, and acts as a calorimeter, capturing the thermal history at these redshifts.
If DM-baryon scattering occurs \textit{after} recombination is complete, it can alter the temperature of baryons at that time. 
The sky-averaged signal and its fluctuations are both sensitive to this effect \cite{Munoz_15, Barkana_18, Fialkov_18, Munoz_18,2018arXiv180210094M}.
Such a late-time scattering scenario arises, for example, in a simple and well-motivated model whereby DM particles carry a small electric charge and exhibit Coulomb-like interactions (``millicharged'' DM) \cite{2017arXiv170704591B,2019arXiv190208623D}.
This scenario is challenging to test in direct detection experiments \cite{2011PhRvD..83f3509M}, and cosmological probes like the 21-cm signal may be an optimal detection channel.
However, while the field of 21-cm cosmology is rapidly advancing, measurement of the 21-cm signal is still difficult to make, and necessitates control of an overwhelming Galactic-foreground systematic uncertainty. 
As was demonstrated with the recent analyses of the EDGES result \cite{Bowman_18}, cross-checks with other probes will likely be essential for the interpretation of future anomalies in the 21-cm signal as new physics \cite{Barkana:2018lgd,Kovetz:2018zan,Berlin:2018sjs}. 

\textbf{Galaxies.}
Galaxies trace structure on even smaller scales, and thus hold a bold promise as a probe of DM physics \cite{2019arXiv190201055D}.
Given the physical scales involved, studies of the abundances of dwarf and satellite galaxies, in particular, could unlock orders of magnitude improvement in sensitivity to DM interactions, compared to the high-redshift and large-scale probes discussed earlier.
However, robust inference of fundamental physics from studies of galaxies will crucially depend on improvements in modeling and numerical simulation of their formation and evolution within cosmologies with interacting DM, and understanding of baryonic effects---a challenge that will need much attention in the coming decade \cite{2019arXiv190201055D}.\footnote{Observables such as galaxy clustering and weak lensing are also relevant for this discussion and could potentially test the same physics; however, they are a prime focus of other science white papers.}

\vspace{-0.3cm}
\subsection{Scattering with Neutrinos}

Neutrinos are one of the least well understood parts of the Standard Model of particle physics; indeed, non-vanishing neutrino masses require new physics to explain. 
In addition, terrestrial neutrino experiments and probes of astrophysical neutrinos have seen a number of anomalies that may indicate new physics \cite{Athanassopoulos:1996jb, Aguilar:2001ty,AguilarArevalo:2007it,2018arXiv180512028M,Gorham:2016zah,Gorham:2018ydl}.
The idea that DM may communicate with visible matter through interactions with neutrinos is an intriguing possibility; however, laboratory tests of this scenario are currently unfeasible.
Meanwhile, the high abundance of cosmological neutrinos allows for observational tests of their possible link with DM.

If neutrinos efficiently scatter with DM seconds after the Big Bang, the DM-neutrino fluid undergoes acoustic oscillations and diffusion damping, and is smoother than ordinary free-streaming neutrinos, with a smaller sound speed.
CMB anisotropy captures this physics through a suppression of power and a shift of the acoustic peaks towards smaller angular scale---all used to probe DM-neutrino scattering with current CMB data \cite{Mangano:2006mp,Escudero:2015yka,DiValentino:2017oaw,Diacoumis:2018ezi}.
The small-scale power in matter fluctuations is also suppressed in this scenario, and can, in principle, probe even weaker interactions than the CMB; the best current bounds on the DM-neutrino scattering cross section come from Lyman-$\alpha$ forest observations \cite{Wilkinson:2014ksa}.
As in the case of DM-baryon scattering, future CMB, Lyman-$\alpha$, and galaxy surveys will increase the sensitivity to DM-neutrino scattering by mapping out structure on even smaller scales.

\vspace{-0.3cm}
\subsection{Annihilation and Decay}

If DM interacts with visible particles, it may also annihilate or decay, further altering the thermal history of the Universe. 
If annihilation or decay products include electromagnetically interacting particles, these particles can generically heat and ionize the baryonic gas.
Cosmological searches have distinct advantages over classic indirect searches for DM annihilation and decay, as they do not suffer from astrophysical backgrounds and large uncertainties in the distribution of DM in the target systems.
Furthermore, they can probe processes with no detectable signals in terrestrial-scale experiments and in the local Universe; for example, decays with lifetimes comparable to the age of the Universe, and decays into invisible channels, such as neutrinos or new dark particles \cite{Poulin:2016nat,Poulin:2016anj}. 

Increasing the ionization fraction near the time of recombination can affect the CMB anisotropy \cite{Adams:1998nr,Chen:2003gz, Padmanabhan:2005es}. 
\textit{Planck} measurements of CMB temperature and polarization anisotropy on degree angular scales provide some of the strongest and most robust bounds on annihilations and decays of sub-GeV DM, complementing indirect searches that probe heavier DM candidates \cite{Aghanim:2018eyx,Slatyer:2016qyl}. 
Measurements from the next-generation ground-based CMB experiments can improve sensitivity to DM annihilation cross section and lifetime by a factor of a few.
Furthermore, as discussed in the context of DM-baryon interactions, future observations of the cosmological 21-cm signal from atomic hydrogen can sensitively track gas temperature during cosmic Dark Ages and Reionization.
As such, they can capture any new energy injection from DM decays and annihilations in the post-recombination Universe, likely surpassing the sensitivity of the CMB anisotropy to the same processes \cite{Furlanetto:2006wp,Valdes:2007cu,Evoli:2014pva,Lopez-Honorez:2016sur,Poulin:2016anj}. 

\vspace{-0.4cm}
\subsection{Interactions with Dark Radiation}

Models in which the dark sector is complex and contains not only dark matter, but also dark radiation (DR) that interacts with DM particles, are motivated in several ways \cite{2016arXiv160808632A}: they generically arise in theories proposed to explain the hierarchy problem of the Standard Model of particles \cite{Arkani-Hamed:2016rle, Chacko:2018vss}; they could explain the anomalously low large-scale amplitude of matter fluctuations seen in some weak-lensing surveys \cite{Lesgourgues:2015wza,Chacko:2016kgg,Buen-Abad:2017gxg,Krall:2017xcw}; and they are a specific case of self-interacting DM, proposed as a possible solution for putative anomalies in DM structure on sub-galactic scales \cite{Tulin:2012wi,Tulin:2013teo,Kaplinghat:2015aga,Bullock:2017xww}. 
Cosmology offers a unique way to probe DM-DR interactions, especially if DM resides in a secluded sector that only weakly couples to known particles.

Similar to how photon pressure prohibits the growth of baryon fluctuations until recombination, DM interacting with DR in the early Universe experiences suppressed growth of structure, as compared to a scenario with no DR. 
The resulting suppression of power is captured in the CMB anisotropy and all tracers of large-scale structure (similar to the case of DM-baryon interactions): galaxy clustering, galaxy weak lensing, the Lyman-$\alpha$ forest, and the cosmological 21-cm signal \cite{Boehm:2001hm,Cyr-Racine:2013fsa,Cyr-Racine:2015ihg}. 
Notably, cosmological signatures of DR are distinct from signatures of relativistic particles that do not couple to DM, such as the standard free-streaming neutrinos~\cite{Bashinsky:2003tk,Follin:2015hya,Baumann:2015rya}.
High-precision measurements of small-angular-scale CMB polarization anisotropy with the next-generation ground-based experiments could robustly detect or rule out DM-DR interactions taking place in the early Universe, even in the case where only a small fraction (less than $5\%$) of DM couples to DR. 
As in previous DM scenarios, other tracers of matter could probe even smaller scales, and weaker interactions. 

\vspace{-0.5cm}
\section{Recommendations}
\label{sec:recommendations}
\vspace{-0.2cm}
The next decade of observations will open avenues to broadly probe the fundamental nature of DM in context of compelling theoretical scenarios, complementing the reach of terrestrial experiments.
Many individual probes---the next-generation CMB and galaxy surveys, and 21-cm line intensity measurements---will reach sufficient raw sensitivity to potentially uncover DM signals that are invisible to the present-day searches.
However, a linchpin to their robustness as probes of fundamental physics will be in their synergies: \textbf{cross-validation between different data sets will be necessary to establish a discovery and measure properties of DM particles with cosmological observations}. 
The following are all central components to realizing the full potential of this program in the coming decade:
\vspace{-0.cm}
\begin{itemize}
    \item \textbf{Observations:}
    \begin{itemize}
        \item Next-generation measurements of high-multipole CMB anisotropy and lensing.
        \item Next-generation surveys of large-scale structure (21-cm tomography, Lyman-$\alpha$ forest, and galaxy surveys).
        \item Thermal-history measurements (global 21-cm signal, CMB spectral distortions).
        \item Near-field cosmology (local probes of small-scale structure).
    \end{itemize}
    \vspace{-0.2cm}
    \item \textbf{Theory and analysis:}
    \begin{itemize}
        \item High-accuracy theoretical prediction of cosmological DM signals.
        \item Simulation and modeling of non-linearities in novel DM cosmologies.
        \item Understanding of baryonic systematic uncertainties in DM-related inference.
        \item Frameworks for joint analyses of cosmological and other DM probes.
        \end{itemize}
\end{itemize}
\pagebreak
\bibliographystyle{ieeetr}
\bibliography{DMCosmology}

\end{document}